\documentclass[useAMS,usenatbib,usegraphicx,letterpaper]{mn2e}

\usepackage{multirow} 
\usepackage{epsfig}
\usepackage{subfigure}
\usepackage{wasysym}

\newcommand{\etal}{et al.}

\newcommand{\swi}{SWIFT~J2127.4+5654 }

\def\xmm{{\it XMM-Newton} }

\def\suzaku{{\it Suzaku} }

\def\mnras{MNRAS}
\def\apj{ApJ}

\def\aap{A\&A}

\def\apjl{ApJ Letters}

\def\xr{X--ray }

\title[The size of the X-ray source in SWIFT~J2127.4+5654] {The
  size of the X--ray emitting region in SWIFT~J2127.4+5654 via
  a broad line region cloud X--ray eclipse}
\author[M. Sanfrutos \etal]
       {M. Sanfrutos$^{1}$\thanks{sanfrutoscm@cab.inta--csic.es},
         G. Miniutti$^{1}$, B. Ag\'is-Gonz\'alez$^{1}$,
         A. C. Fabian$^{2}$, J. M. Miller$^{3}$, \newauthor
         F. Panessa$^{4}$ and A. Zoghbi$^{5,6}$ \\ \\ $^1$ Centro de
         Astrobiolog\'ia (CSIC--INTA), Dep. de Astrof\'isica; ESA,
         P.O: Box 78, E-28691, Villanueva de la Ca\~nada, Madrid,
         Spain\\ $^2$ Institute of Astronomy, Madingley Road,
         Cambridge CB3 0HA\\ $^3$ Department of Astronomy, University
         of Michigan, 500 Church St., Ann Arbor, MI48109, USA\\ $^4$
         INAF - Istituto di Astrofisica e Planetologia Spaziali di
         Roma (IAPS), Via del Fosso del Cavaliere 100, 00133 Roma,
         Italy\\ $^5$ Department of Astronomy, University of Maryland,
         College Park, MD 20742-2421, USA\\ $^6$ Joint Space-Science
         Institute (JSI), College Park, MD 20742-2421, USA\\ }

\pagerange{\pageref{firstpage}--\pageref{lastpage}}
\pubyear{2001}

\usepackage{times}

\begin{document}

\label{firstpage}

\maketitle

\begin{abstract}
We present results obtained from the time--resolved X--ray spectral
analysis of the Narrow--Line--Seyfert~1 galaxy SWIFT~J2127.4+5654
during a $\sim 130$~ks {\it XMM--Newton} observation. We reveal large
spectral variations, especially during the first $\sim$~90~ks of the
{\it XMM--Newton} exposure. The spectral variability can be attributed
to a partial eclipse of the X-ray source by an intervening
low--ionization/cold absorbing structure (cloud) with column density
$N_{\rm{H}} = 2.0^{+0.2}_{-0.3}\times 10^{22}$~cm$^{-2}$ which
gradually covers and then uncovers the X--ray emitting region with
covering fraction ranging from zero to $\sim 43$~per cent. Our
analysis enables us to constrain the size, number density, and
location of the absorbing cloud with good accuracy. We infer a cloud
size (diameter) of $D_{\rm c} \leq 1.5 \times 10^{13}$~cm,
corresponding to a density of $n_{\rm c}\geq 1.5\times 10^9$~cm$^{-3}$
at a distance of $R_{\rm c} \geq 4.3\times 10^{16}$~cm from the
central black hole. All of the inferred quantities concur to identify
the absorbing structure with one single cloud associated with the
broad line region of SWIFT~J2127.4+5654. We are also able to constrain
the X--ray emitting region size (diameter) to be $D_{\rm s} \leq
2.3\times 10^{13}$~cm which, assuming the black hole mass estimated
from single--epoch optical spectroscopy ($1.5\times 10^7~M_\odot$),
translates into $D_{\rm s} \leq 10.5$ gravitational radii ($r_g$) with
larger sizes (in $r_g$) being associated with smaller black hole
masses, and viceversa. We also confirm the presence of a
relativistically distorted reflection component off the inner
accretion disc giving rise to a broad relatvistic Fe K emission line
and small soft excess (small because of the high Galactic column density),
supporting the measurement of an intermediate black hole spin in
SWIFT~J2127.4+5654 that was obtained from a previous {\it Suzaku}
observation. 
\end{abstract}


\section{Introduction}

\xr flux and spectral variability is a rather common property of
Active Galactic Nuclei (AGN). Spectral variability on relatively long
timescales (months to years) is often associated with absorption
variability (e.g. Warwick et al. 1988; Risaliti et al. 2002; Marinucci
et al. 2012). In recent years, several examples of short--timescale
(hours to days) absorption variability have been reported
(e.g. NGC~4388, Elvis et al. 2004; NGC~4151, Puccetti et al. 2007;
NGC~1365, Risaliti et al. 2005; 2007; 2009 and Maiolino et al. 2010;
NGC~7582, Bianchi et al. 2009; Mrk~766).  The
detailed analysis of the short--timescale absorption variability in
these sources strongly suggests the presence of absorbing structures
(clouds) crossing the line--of--sight to a rather compact X--ray
source. The data are generally consistent with the presence of
an ensemble of compact, cold clouds with typical column density of
$10^{22}-10^{24}$~cm$^{-2}$, density of $10^9-10^{11}$~cm$^{-3}$,
velocity of few times $10^{3}$~km~s$^{-1}$, and distance of
$10^2-10^4~r_g$ from the centre. These properties are remarkably
similar to those of the clouds responsible for the emission of the
broad optical and UV emission lines in AGN. Hence, X--ray absorption
variability studies suggest to identify the \xr compact absorbers responsible
for the shortest timescale absorption variability events with broad
line region (BLR) clouds, while variability on longer timescales
(months to years) is likely associated with more extended structures,
possibly associated with a clumpy torus (Nenkova et al. 2008). Here we
report results from a $\sim 130$~ks observation of the Narrow Line
Seyfert~1 galaxy SWIFT~J2127.4+5654 with {\it XMM--Newton}.


\swi (a.k.a. IGR~J21277+5656) is a low Galactic latitude hard X--ray
source which has been first detected at hard X--rays with
{\it{Swift}}/BAT (Tueller et al. 2005). Later, the source was
identified as a Narrow--Line Seyfert 1 (NLS1) galaxy at redshift
0.0147 based on the observed H$\alpha$ FWHM of $\sim$~1180~km/s
(Halpern 2006). Subsequent work by Malizia et al. (2008) supported the
NLS1 classification by obtaining, despite a slightly larger FWHM of
$\sim$~2000~km/s, a relatively low [{O\,\textsc{iii}}]/H$_\beta$ ratio
of $0.72\pm 0.05$ and significantly enhanced {Fe\,\textsc{ii}}
emission ({Fe\,\textsc{ii}}/H$_\beta =1.3\pm 0.2$). \swi was then
observed with \suzaku for a total net exposure of 92~ks and results
have been presented and discussed by Miniutti et al. (2009). The major
result of the \suzaku observation is the detection of a 
relativistically broadened Fe K$\alpha$ emission line which strongly
suggests that \swi is powered by accretion onto a rotating Kerr black
hole, with an intermediate spin value of $a=0.6\pm 0.2$. This result
has been confirmed (using the same data set) also by Patrick et
al. (2010) who report $a= 0.70^{+0.10}_{-0.14}$, although a different
interpretation of the Fe~K$\alpha$ shape, based on reprocessing and
scattering in a Compton--thick disc--wind rather than the disc itself, has also been
proposed (Tatum et al. 2012).

\section{X-Ray Observations}

\xmm observed \swi on 2010 November 29 for a full revolution
($\sim$~130~ks). The observation (0655450101) was performed using the
Small Window mode for all EPIC cameras, with the optical thin filter
applied. The data were reduced as standard using the dedicated
{\small{SAS v11.0}} software. Observation--dependent EPIC and RGS
response files were generated using the {\small{RMFGEN}} and
{\small{ARFGEN}} tasks. EPIC source products were extracted from
circular regions of 40$''$ centred on the source, and the
corresponding background was estimated by using source--free nearby
regions.  In the time--resolved analysis we present in this work, each
time--interval is associated with its own background spectrum
contributing less than 2~per cent in any of the considered
time--intervals. After filtering for high background periods which
occur at the beginning and end of the exposure, the net exposure is
$\sim$~84~ks in the EPIC--pn spectrum and $\sim$~109~ks and
$\sim$~110~ks in the MOS~1 and MOS~2 spectra respectively. All the
EPIC spectra are extracted with common good--time--intervals and they
are all grouped in order to guarantee that each background--subtracted
bin has at least 25 counts.

\begin{figure}
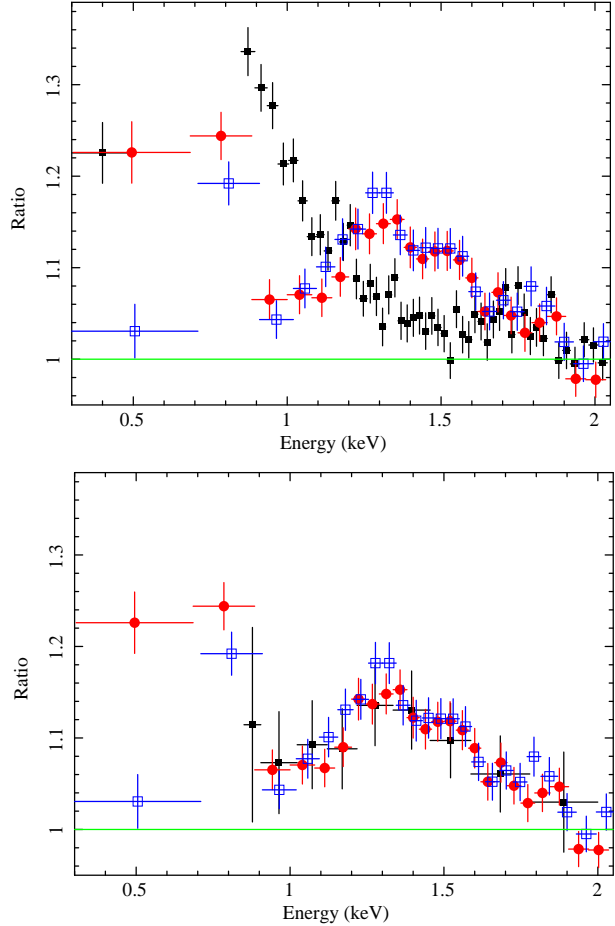

\begin{center}
\includegraphics[width=0.34\textwidth,height=0.45\textwidth,angle=-90]{pnmos.ps}
{\vspace{0.2cm}}
\includegraphics[width=0.34\textwidth,height=0.45\textwidth,angle=-90]{ratio2.ps}
\caption{\label{discrep} In the upper panel, we show a comparison
  between the EPIC pn (filled black squares), MOS~1 (filled circles,
  red in the on--line version), and MOS~2 (empty squares, blue in the
  on--line version) spectra in the soft X--ray band. All spectra are
  fitted in the 2--4~keV and 7.5--10~keV band (i.e. ignoring the Fe K
  band) with a common spectral model comprising Galactic absorption
  and a power law. See the upper panel of Fig.~\ref{models} for a
  comparison between the pn and MOS data at higher energies using the
  same spectral model used here. In the lower panel, we show the same
  MOS data as above, but we plot the (heavily rebinned) RGS~2
  data--to--model ratio instead of the EPIC pn (we omit the RGS~1 data
  here because of the gap around the crucial 0.9-1.1~keV band
  affecting the RGS~1 detector). The RGS data below 0.8--0.9~keV are
  too noisy to be used for this comparison.}
\end{center}
\end{figure}

The MOS and pn spectra are consistent with each other above
$\sim$~1.6~keV. However, as shown in the upper panel of
Fig.~\ref{discrep}, the spectra are not consistent with each other at
softer energies. The largest discrepancies are between the pn and the
two MOS detectors below 1.6~keV and between the two MOS cameras below
0.9~keV. In the lower panel, we plot the RGS~2 data--to--model ratio
instead of the EPIC pn. The RGS spectrum confirms the reliability of
the MOS data above 0.9 ~keV and suggests to ignore the pn softest
data. Moreover, the spectral shape in the previous {\it Suzaku}
observation (Miniutti et al. 2009) agrees much better with the MOS/RGS
spectra than with the pn one. This comparative analysis suggests to
make a conservative choice of the energy ranges to be used for
spectral analysis. In our work we consider the MOS data in the
0.9--10~keV band, and the pn data in the 1.6--10~keV band. Our choice
of reliable energy bands to be used for the spectral analysis is in
fact very conservative, and we have checked a--posteriori that almost
exactly the same results on the most relevant best--fitting parameters
are obtained considering all data in the 0.5--10~keV band, although
with worse statistical results due to the discrepancies between the pn
and the MOS spectra. In any case, it should also be mentioned that,
since the low Galactic latitude of \swi corresponds to a relatively
high Galactic column density ($7.65\times 10^{21}$~cm$^{-2}$, Kalberla
et al. 2005), soft X--ray data below 1~keV only provide $\sim$~7 per
cent of the total collected counts, with very little impact on the
overall analysis.  When needed to convert fluxes into luminosities, we
adopt a cosmology with H$_0=70$~km~s$^{-1}$~Mpc$^{-1}$,
$\Omega_\Lambda= 0.73$, and $\Omega_M = 0.27$.

\begin{figure}
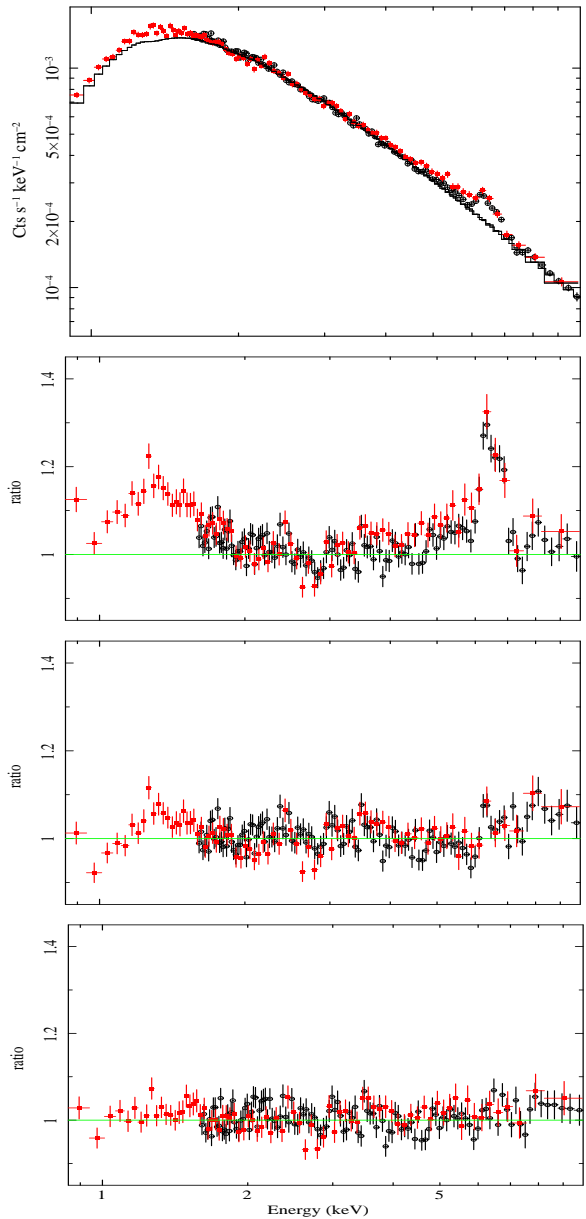

\begin{center}
\includegraphics[width=0.25\textwidth,height=0.43\textwidth,angle=-90]{data1.ps}
{\vspace{0.2cm}}
\includegraphics[width=0.2\textwidth,height=0.43\textwidth,angle=-90]{r0.ps}
{\vspace{0.2cm}}
\includegraphics[width=0.2\textwidth,height=0.43\textwidth,angle=-90]{r1.ps}
{\vspace{0.2cm}}
\includegraphics[width=0.22\textwidth,height=0.43\textwidth,angle=-90]{r2.ps}
\caption{In the upper panel, we show the EPIC pn (empty circles) and
  MOS ~1 data (filled squares, red in the on--line version) divided by
  the detector effective area to show the real spectral shape,
  together with the best--fitting power law model fitted in the
  2--4~keV and 7.5--10~keV data.  Although the MOS~2 data were also
  used in the spectral analysis in the 0.9--10~keV band, they are
  omitted here for visual clairty. In the second panel, we show the
  resulting data--to--model ratio. In the third panel, we show the
  data--to--model ratio resulting from including a relativistacally
  blurred Fe emission line. In the lower panel we show the final
  best--fitting model comprising a relativistically blurred reflection
  model plus a series of narrow emission lines as detailed in the
  text.}
\label{models}
\end{center}
\end{figure}

\section{The time--averaged spectrum}
\label{timeav}

We start our analysis by considering the time--averaged X--ray
spectrum of \swi in the 0.9--10~keV band using all EPIC detectors. In
the upper two panels of Fig.~\ref{models} we show the data--to--model
ratio resulting from a simple absorbed power law model fitted in the
2--4~keV and 7.5--10~keV, i.e. ignoring the Fe K band (as mentioned,
we use the MOS data in the 0.9--10~keV band, and the pn data in the
1.6--10~keV band). Absorption is here modelled with only the Galactic
column density of $7.65\times 10^{21}$~cm$^{-2}$ (Kalberla et
al. 2005). The continuum is modelled with a power law of
$1.81\pm0.02$.  The model leaves significant residuals around the Fe K
band, where a broad feature reminiscent of a relativistic broad Fe
line is seen, and below $\sim 2$~keV where a soft X--ray excess is
detected.

We then add a relativistic Fe line using the the {\small KERRCONV}
relativistic convolution model from Brenneman \& Reynolds (2006)
applied to a 6.4~keV Gaussian emission line with zero intrinsic
width. The statistical description of the data is relatively good with
$\chi^2 = 4080$ for 3838 degrees of freedom (dof). Adding an intrinsic
neutral absorber at the redshift of the source improves the statistic
to $\chi^2 = 4030$ for 3837 dof. The nature and properties of the
absorber will be discussed in detail in subsequent Sections. Here we
only mention that we obtain a column density of $\sim 2\times
10^{21}$~cm$^{-2}$ in addition to the Galactic one. The resulting
data--to--model ratio is shown in the third panel of
Fig.~\ref{models}. Some residuals are still present in the soft and
hard bands.

As the detected broad line should be associated with a disc reflection
continuum, we replace the relativistic line model with a full disc
reflection model from partially ionized gas (Ross \& Fabian 2005) to
which we apply the same relativistic convolution model as before. The
model describes the data better with $3976$ for 3836 dof, and accounts
for both the soft and hard residuals seen in the middle panel of
Fig.~\ref{models}. 

Further residuals are however left around the Fe K band, suggesting
the presence of a series of narrow emission lines. We find that four
narrow lines are indeed required by the data. Their energy is
consistent with 6.4~keV (Fe K$\alpha$), 6.7~keV (Fe~\textsc{xxv}
Ly$\alpha$), 6.97~keV (Fe~\textsc{xxvi} Ly$\alpha$), and 8.25~keV
(Fe~\textsc{xxvi} Ly$\beta$) so that we fix their energy at the
theoretical values. The statistical improvement for the inclusion of
these additional narrow lines is $\Delta\chi^2_{6.4} = -22$,
$\Delta\chi^2_{6.7} = -11$, $\Delta\chi^2_{6.97} = -10$, and
$\Delta\chi^2_{8.25} = -7$, where each line contributes with one
additional degre of freedom only (its normalization). Only the narrow
Fe K$\alpha$ emission line was detected during the previous {\it
  Suzaku} observation. However, the 2--10~keV flux was a factor of
$\sim 2$ higher than during the {\it XMM--Newton} observation, which
can easily explain the non--detection of the ionized Fe lines as, for
constant--intensity lines, their EW drops by a similar factor and
becomes consistent with zero within the uncertainties. This likely
signals that the ionized lines are associated with emission from
extended photo--ionized gas, so that their flux remains constant and
the lines can only be detected when the X--ray continuum flux is
relatively low, as during the {\it XMM--Newton} observation. The same
applies to the narrow Fe K$\alpha$ line at 6.4~keV, as its EW during
the {\it Suzaku} observation ($13\pm 9$~eV) is broadly consistent with
being half of that measured here ($\sim 50\pm 17$~eV), which is what
is expected for a line with a constant intensity and a continuum
variation by a factor of $\sim 2$. We point out that if the width of
the Fe K$\alpha$ line is allowed to vary during the fit, we have
$\sigma \leq 0.1$~keV, corresponding to a FWHM$\leq 1.1\times
10^4$~km~s$^{-1}$ (consistent with any production site, from the
innermost BLR outwards). With the addition of these narrow emission
lines, the final best--fitting model for the 0.9--10~keV spectrum
produces a statistical result of $3926$ for 3832 dof. The
best--fitting data--to--model ratio is shown in the lower panel of
Fig.~\ref{models}. The most important best--fitting parameters are
reported in Table~\ref{tableaveraged}. All relativistic parameters
(emissivity index, black hole spin, disc inclination) are consistent
with the results of Miniutti et al. (2009) who measured $q\sim 5.3$,
$a\sim 0.6$ and $i\sim 46^\circ$ from the previous {\it Suzaku}
observation. The good agreement between the {\it XMM--Newton} and {\it
  Suzaku} data provides independent support to the interpretation of
the spectral shape in terms of relativistically distorted disc
reflection, as originally proposed by Miniutti et al. (2009).

\begin{table}
\caption{Best--fitting parameters from the time--averaged 0.9--10~keV
  spectral analysis of \swi. The column density of the intrinsic absorber at the
  redshift of the source is in units of $10^{22}$~cm$^{-2}$. The
  reflector ionization state is given in units of erg~cm~s$^{-1}$. The
  2--10~keV flux is in units of $10^{-11}$~erg~cm$^{-2}$~s$^{-1}$ and
  is as observed, while the 2--10~keV luminosity is unabsorbed and
  expressed in $10^{42}$~erg~s$^{-1}$. The equivalent widths (EW) are
  expressed in units of eV and they refer to the narrow emission lines at
  6.4~keV (Fe K$\alpha$), 6.7~keV (Fe~\small{xxv} Ly$\alpha$),
  6.97~keV (Fe~\small{xxvi} Ly$\alpha$), and 8.25~keV
  (Fe~\small{xxvi} Ly$\beta$).}
\label{tableaveraged}      
\begin{center}
\begin{tabular}{l c c }
\hline\hline                 
$\Gamma$ & .................................................. & $2.00\pm 0.04$\\
$N_{\rm H}$& .................................................. &$0.2\pm 0.1$ \\
\hline
$q$ & .................................................. &$4.9\pm 0.9$\\
$a$ & .................................................. &$0.5 \pm 0.3$\\
$i$ & .................................................. &$44^\circ \pm 8^\circ$\\
$\xi_{\rm{ref}}$ & .................................................. &$10\pm 6$\\
\hline
$EW_{\rm{6.4}}$ & .................................................. &$50\pm 17$ \\
$EW_{\rm{6.7}}$ & .................................................. &$23\pm 11$ \\
$EW_{\rm{6.97}}$ & .................................................. &$25\pm 10$\\
$EW_{\rm{8.25}}$ & .................................................. &$18\pm 14$\\
\hline
$F_{2-10}$ & .................................................. &$1.90\pm 0.04$\\
$L_{2-10}$ & .................................................. &$9.6\pm 0.2$\\
\hline
$\chi^2/{\rm{dof}}$ & .................................................. &3926/3832\\
\hline
\hline
\end{tabular}
\end{center}
\end{table}

\section{Spectral variability}

\begin{figure}
\begin{center}
\includegraphics[width=0.34\textwidth,height=0.45\textwidth,angle=-90]{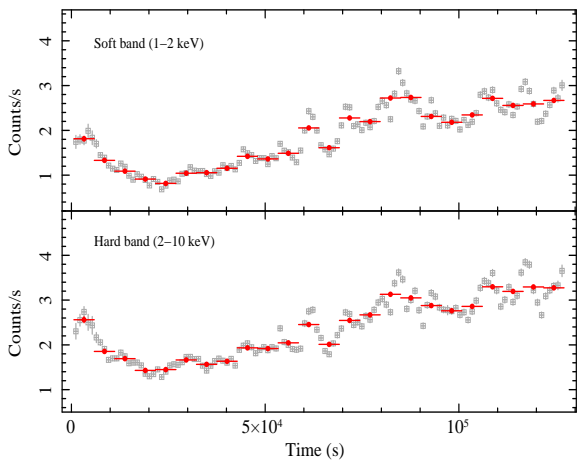}
\caption{\label{hi_lo} The soft (S: 1--2~keV, upper) and hard (H:
  2--10~keV, lower) X--ray light curves of \swi from the pn detector
  are shown in bins of 500~s (empty grey squares) and 5271.8~s (filled
  red circles). The latter bin--size is chosen to divide the total
  exposure in 24 uniform time--intervals which will be used to perform
  a time--resolved spectral analysis of the data.}
\end{center}
\end{figure}

\begin{figure}
\begin{center}
\includegraphics[width=0.34\textwidth,height=0.45\textwidth,angle=-90]{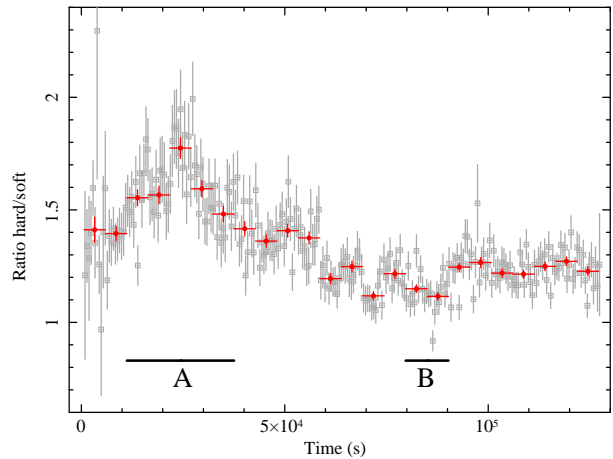}
\caption{\label{hr} The hard--to--soft ratio (H/S) light curve of \swi
  from the pn detector is shown with the same bin--size as in
  Fig.~\ref{hi_lo}. \swi exhibits highly significant spectral
  variability during the first $\sim$~90~ks, while H/S remains
  approximately constant during the subsequent $\sim$~35~ks. We also
  highlight the two intervals of highest (A) and lowest (B) H/S that
  are used in the spectral variability analysis (see text for
  details).}
\end{center}
\end{figure}

As already shown in previous work (e.g. Miniutti et al. 2009), \swi is
highly variable in X--rays, an almost ubiquitous property of NLS1
galaxies. X--ray flux variability is also present in the {\it
  XMM--Newton} observation. In Fig.~\ref{hi_lo}, we show the EPIC--pn
light curve in a soft (S: 1--2~keV) and a hard (H: 2--10~keV)
band\footnote{We extend the EPIC pn energy range down to 1~keV
  although we only use the pn data above 1.6~keV in our spectral
  analysis, as light curves and, even more so, hardness ratios are reasonably
  calibration--independent. The shape of the pn light curves shown in
  Fig.~\ref{hi_lo} and Fig.~\ref{hr} has been checked against the MOS
  data which confirm that the EPIC pn light curves are reliable.}. The
source is variable on all probed timescales and exhibits variability
up to a factor of $\sim$~4 during the {\it XMM--Newton}
exposure. Significant bin--to--bin variability is present down to
the chosen bin--size of 500~s, with more than one third of any two
consecutive bins being inconsistent with each other. We then consider
the hard--to--soft ratio (H/S) light curve, to investigate whether
spectral variability is also present during the observation. The H/S
light curve is shown in Fig.~\ref{hr} with the same bin--size as in
Fig.~\ref{hi_lo}. Clear spectral variability is present in the first
$\sim$~90~ks of the exposure, while H/S remains approximately stable
in the subsequent $\sim$~35~ks.

\subsection{Time--resolved analysis and spectral variability}

\begin{figure}
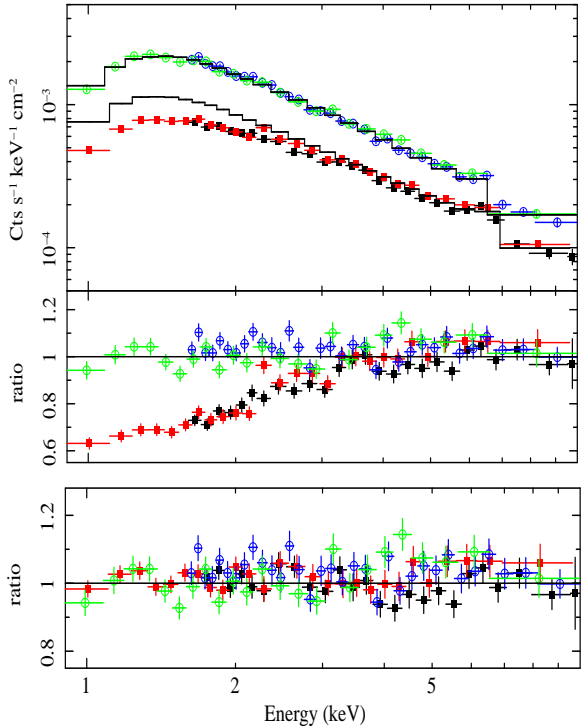

\begin{center}
\includegraphics[width=0.35\textwidth,height=0.43\textwidth,angle=-90]{extr1.ps}
{\vspace{0.2cm}}
\includegraphics[width=0.18\textwidth,height=0.43\textwidth,angle=-90]{extr2.ps}
\caption{In the upper panel we show the data divided by the
  appropriate detector effective area, best--fitting models, and
  data--to--model ratio for spectra A (lower two spectra, filled
  squares, black and red in the on--line version for the pn and MOS~1
  cameras respectively) and B (upper two spectra, empty circles, blue
  and green in the on--line version for the pn and MOS~1 cameras
  respectively). The MOS~2 data are also used in the analysis in the
  0.9--10~keV band (as for the MOS~1 data) but are not shown here for
  visual clarity. The spectral model is fitted only above 3.5~keV and
  then extrapolated to soft energies to show the significant spectral
  difference between the two time--intervals. In the lower panel, we
  show the final best--fitting data--to--model ratio where the spectra
  A residuals are accounted for by an additional neutral absorber only
  partially covering the X--ray continuum (see text for details).}
\label{extr}
\end{center}
\end{figure}

In order to understand the origin of the spectral variability, we
first extract X--ray spectra from the two time--intervals, denoted as
A and B in Fig.~\ref{hr}, which correspond to the highest and lowest
H/S ( $1.59^{+0.24}_{-0.15}$ and $1.13\pm 0.04$ respectively). The
different duration of intervals A and B is justified by the different
flux levels, and by the requirement that the two spectra share a
similar spectral quality (i.e that the spectra have similar number of
counts). Interval A is $\sim$~26~ks long and corresponds to a
0.9--10~keV flux of $\sim 1.46\times 10^{-11}$~erg~cm$^{-2}$~s$^{-1}$, while
interval B is $\sim$~11~ks long with a flux of $\sim 2.83\times
10^{-11}$~erg~cm$^{-2}$~s$^{-1}$ in the same band. Visual inspection
of spectra A and B shows that the two spectra only differ spectrally
below $\sim$~3--4~keV where spectrum A has a deficit of soft photons,
possibly the signature of extra--absorption with respect to the lower
H/S spectrum B.

The spectra from intervals A and B are fitted jointly with the
best--fitting model inferred from the time--avergaed spectral
analysis. We fix the relativistic parameters to their time--averaged
best--fitting values, as the lower quality of the time--resolved
spectra does not allow us to better constrain them. We also fix the
normalization of all the narrow emission lines to their time--averaged
best--fitting values as these components are not expected to vary on
short timescales. We start our analysis by considering only the data
above 3.5~keV. We initially allow for possible variation in the
continuum photon index $\Gamma$. However, as $\Gamma$ turns out to be
consistent with being constant, we force it to be the same in
the A and B spectra. We reach an excellent description of the hard
X--ray spectra of intervals A and B at energies above 3.5~keV with
$\chi^2= 1190$ for 1196 dof. However, the extrapolation to soft
energies reveals a large spectral difference between the two
intervals, as shown in the upper panel of Fig.~\ref{extr}. The
spectrum from interval A appear to be significantly more absorbed than
that from interval B, as already suggested by the H/S difference
between the two time--intervals in Fig.~\ref{hr}.

We then add a layer of absorbing gas to our model. We use a partial
covering model for neutral gas (the {\small{ZPCFABS}} model). In
principle, both the column density and the covering fraction may be
variable. However, due to the relatively low quality of the data, the
two parameters cannot be constrained independently. In order to
explain physically the short--timescale spectral variability, we
assume that the absorber column density is the same in both spectrum A
and B, while its covering fraction is allowed to vary
independently. The model provides an excellent description of the data
($\chi^2= 3179$ for 3135 dof). In the lower panel of Fig.~\ref{extr}
we show the resulting best--fitting data--to--model ratio for the
spectra A and B. The observed variability between intervals A and B
can be entirely explained by a layer of neutral absorbing gas with
column density $N_{\rm{H}} = 2.0^{+0.6}_{-0.5}\times
10^{22}$~cm$^{-2}$ covering $\sim 35$~per cent of the X--ray source during
interval A and $\leq 4$~per cent during interval B. If the photon indices of
spectra A and B are now allowed to vary independently, no further
statistical improvement is obtained, and the two spectral slopes are
consistent with each other.

\subsection{A partial eclipse}

We then consider the spectral variability of \swi on shorter
timescales within the framework of the partial covering model. We
extract 24 spectra (for each of the EPIC cameras) from intervals of
equal duration (i.e. from the 5271.8~s long intervals shown in
Fig.~\ref{hr}), and we apply the same model discussed above to all
spectra. We force all parameters to be the same, except the power law
normalisation, the reflector ionization\footnote{We impose the same
  reflection normalization in all spectra. This is because, by
  definition, two reflection models with the same normalization and
  illuminating power law slope but with different ionization
  parameters satisfy a linear relation between ionization and total
  reflection flux, see e.g. Miniutti et al. 2012. Imposing that this
  is the case by forcing all reflection normalization to be the same
  increases the overall self--consistency of the model, as the two
  quantities are both linearly correlated with the flux irradiating
  the disc. It also helps lowering the number of free parameters in
  the model, reducing the risk of over--modelling data of relatively
  poor quality.}, and the absorber covering fraction. The overall fit
is excellent (reduced $\chi^2 = 0.98$), and we obtain a common
absorber column density of $N_{\rm{H}} = 2.0^{+0.2}_{-0.3}\times
10^{22}$~cm$^{-2}$, consistent with that derived above from the
analysis of intervals A and B.

\begin{figure}
\begin{center}
\includegraphics[width=0.33\textwidth,height=0.43\textwidth,angle=-90]{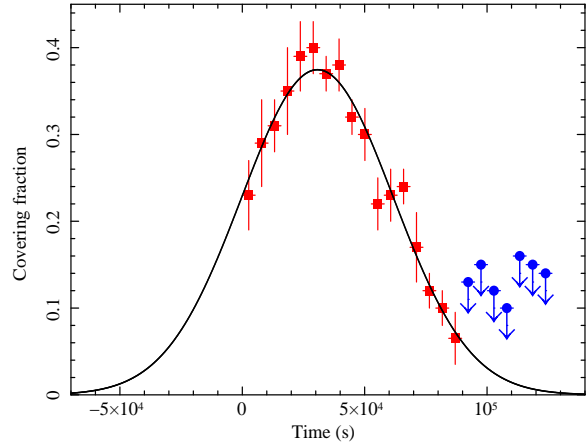}
\caption{\label{cf} The covering fraction ($CF$) evolution
  during the {\it XMM--Newton} observation. Significant $CF$
  variability is present in the first $\sim$~90~ks. Our results
  indicate that the X--ray continuum is partially absorbed during the
  first 90~ks of the exposure by a column density of $\sim 2\times
  10^{22}$~cm$^{-2}$ with variable $CF$, while the data are consistent
  with the X--ray continuum being unabsorbed thereafter. The solid
  curve represents a simple Gaussian model for the $CF$ evolution during
  the first $\sim$~90~ks.}
\end{center}
\end{figure}

The covering fraction ($CF$) during the different intervals is
clearly variable, and its evolution is shown in Fig.~\ref{cf}. The
evolution during the first 90~ks of the exposure is smooth and
can be roughly described by a Gaussian--like shape, while \swi is
consistent with being unabsorbed thereafter. The
Gaussian--like shape of the covering fraction variability suggests
that one single absorbing structure (cloud) with size similar to that
of the X--ray source has crossed the line--of--sight during the first
90~ks of the observation. Below, we discuss the implications of the
main event that characterises the first 90~ks of the observation, and
we derive the properties of the absorber as well as the size of the
X--ray emitting region. Here we simply point out that the
intrinsic absorber detected in the time--averaged spectrum (see
section~\ref{timeav}) is just the flux--weighted average
of the absorption column density and $CF$ during the whole
exposure, so that little physical meaning should be attached to the
derived column density in the time--averaged spectral analysis. 

\section{Cloud properties and the X--ray emitting region size}

The behaviour of the covering fraction data in Fig.~\ref{cf},
identifies three phases during the eclipse: a fast rise of CF from
zero to maximal CF ($\sim 35-45$ per cent), a flat plateau with
maximal CF, and finally a decaying phase from maximal CF to zero.  The
existence of a plateau with maximal CF smaller than 1, as well as the
steepness of the rising and decaying phases, point towards a physical
situation characterised by a cloud, smaller than the source, crossing
our line--of--sight. The other possible scenario, specifically a
larger cloud misaligned with the source would imply a much smoother
and slower growth and decrease of the CF, which is not taking place
here. 

\begin{figure}
\begin{center}
\includegraphics[width=0.45\textwidth,height=0.27\textwidth,angle=0]{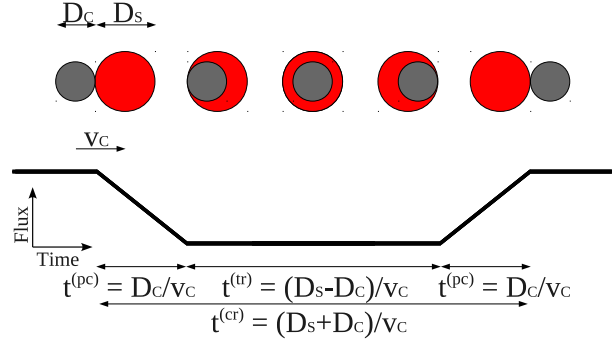}
\caption{\label{phy}The envisaged system geometry. The absorbing cloud
  (grey) moves with velocity $v_{\rm c}$ and partially covers the
  X--ray source (red in the on--line version). The relevant quantities
  used in our discussion ($D_{\rm{c}}$, $D_{\rm{s}}$, $t^{\rm{(pc)}}$, $t^{\rm{(tr)}}$,
  and $t^{\rm{(cr)}}$) are also defined.  }
\end{center}
\end{figure}

We then assume here the simplest possible geometry that can explain
our data, namely the presence of an X--ray source of size (diameter)
$D_{\rm{s}}$ and of an obscuring cloud of size $D_{\rm{c}} \leq
D_{\rm{s}}$, see Fig. \ref{phy}. The $CF$ evolution identifies three
different timescales: (i) the crossing time $t^{\rm{(cr)}} = 160\pm
30$~ks, i.e. the total time during which $CF \neq 0$; (ii) the transit
time $t^{\rm{(tr)}}\leq 26.5$~ks during which the source is maximally
covered (i.e. up to 5 intervals are consistent with being maximally
covered); (iii) the partially covered rising/decaying time
$t^{\rm(pc)} = (t^{\rm{(cr)}} - t^{\rm{(tr)}})/2 =73.5\pm 21.5$~ks
needed to reach the maximal covering fraction starting from zero (and
viceversa). On the other hand all these timescales can be related to
the cloud velocity $v_{\rm c}$ and cloud/source sizes by $v_{\rm c} =
D_{\rm c} /t^{\rm{(pc)}} = (D_{\rm s} - D_{\rm{c}})/t^{\rm{(tr)}} =
(D_{\rm s} + D_{\rm c}) /t^{\rm{(cr)}}$.

{\bf Cloud/source relative sizes ---} Since the maximum covering
fraction $CF_{\rm{max}} \leq 1$, the cloud never completely covers
the X--ray source (or it may, but only for a time shorter than our bin
size of $\sim 5.3$~ks). The simplest interpretation is that
$D_{\rm{c}} = \alpha D_{\rm{s}}$ with $\alpha \leq 1$. There are two
different ways to derive $\alpha$. Firstly, as $CF_{\rm{max}} \geq
0.35$, one has $\alpha = CF_{\rm{max}}^{1/2} \geq 0.59 $. Secondly,
as $(D_{\rm s} - D_{\rm{c}})/t^{\rm{(tr)}} = (D_{\rm s} + D_{\rm c})
/t^{\rm{(cr)}}$, one has $\alpha =
(t^{\rm{(cr)}}-t^{\rm{(tr)}})/(t^{\rm{(cr)}}+t^{\rm{(tr)}}) =
[0.66-1]$. As the two $\alpha$ must be consistent with each other, we
then have that $\alpha=[0.66-1]$.

{\bf Cloud density/distance ---} The combination $n_{\rm c} R_{\rm
  c}^2$ between cloud density and distance can be estimated from the
cloud ionization state $\xi_{\rm c}= L_{\rm{ion}}/(n_{\rm c} R_{\rm
  c}^2)$ if the overall ionising luminosity $L_{\rm{ion}}$ between 1
and 1000~Ry is known. In the case of \swi we have $L_{\rm{ion}} \simeq
2\times 10^{44}$~erg~s$^{-1}$. We have then repeated our spectral
analysis using an ionized absorber (the {\small{ZXIPCF}} model), and
we infer that $\xi_{\rm c} \leq 20$~erg~cm~s$^{-1}$ with similar
column density as inferred for the neutral model, and with consistent
covering fraction evolution (no ionization variability is
detected). Hence, we have $n_{\rm c} R_{\rm c}^2 \geq
10^{43}$~cm$^{-1}$.

We can now use the only relation that has not yet been used, namely
$v_{\rm c} = D_{\rm c} /t^{\rm{(pc)}}$. Assuming that the cloud
transverse velocity is dominated by gravity, and using the
relationship between column density, size, and number density, we then
have $(GM_{\rm{BH}}/R_{\rm c})^{1/2} = D_{\rm c} /t^{\rm{(pc)}} =
N_{\rm H} /(n_{\rm c}t^{\rm{(pc)}})$. For a black hole mass of
$M_{\rm{BH}} = 1.5\times 10^7~M_\odot$ (as derived from single--epoch
optical spectra, see Malizia et al. 2008), $R_{\rm c}
n_{\rm c}^{-2} = [1.9-3.7] \times 10^{-2}$~cm$^{7}$. By combining this
result with $n_{\rm c} R_{\rm c}^2 \geq 10^{43}$~cm$^{-1}$ derived
above, one has $n_{\rm c}^5\geq 7.3 \times 10^{45}$~cm$^{-15}$, which
then gives $n_{\rm c} \geq 1.5\times 10^9$~cm$^{-3}$ and $R_{\rm c}
\geq 4.3\times 10^{16}$~cm.

{\bf Cloud and X--ray source sizes ---} The system is now closed, and
limits on the cloud and X--ray source sizes can be derived. As $D_{\rm
  c} = N_{\rm H}/n_{\rm c}$ one has $D_{\rm c} \leq 1.5\times
10^{13}~{\rm{cm}} = 7~r_g$, where we have used a black hole mass of
$1.5\times 10^7~M_\odot$ to define $1~r_g = 1~GM_{\rm BH}/c^2 =
2.2\times 10^{12}$~cm. The upper limit on the cloud size translates
into an upper limit on the X--ray emitting region size $D_{\rm s} =
D_{\rm c}/\alpha \leq 2.3\times 10^{13}~{\rm{cm}} = 10.5~r_g$.

\section{Summary and conclusions}

We report results from a $\sim$~130~ks observation of the NLS1 galaxy
\swi with {\it XMM--Newton}. We confirm the detection of a
relativistically broadened Fe K$\alpha$ line, originally detected in a
previous {\it Suzaku} observation (Miniutti et al. 2009). All
relativistic parameters are consistent with those inferred from the
{\it Suzaku} data. In particular, we confirm an intermediate black
hole spin of $a\sim 0.5$ and a relatively high observer inclination of
$i\sim 44^\circ$. 

The source exhibits significant flux variability throughout the
observation and, most importantly, spectral variability is also
present during the first 90~ks of the {\it XMM--Newton} exposure. We
perform a time--resolved spectral analysis of the {\it XMM--Newton}
data with the goal of investigating the origin of the observed
spectral variability. Our results are consistent with the same
baseline spectral model as in the previous {\it Suzaku} observation
(e.g. Miniutti et al. 2009) affected by additional neutral absorption
only partially--covering the X--ray source during the first 90~ks of
the {\it XMM--Newton} exposure. Assuming that the spectral variability
is driven by changes of the absorber CF, we show that the CF evolution
is consistent with one single absorbing structure (cloud) crossing our
line--of--sight during the first 90~ks of the exposure.

By considering the first 90~ks of the observation within the framework
of a simple source--cloud geometry, we obtain the following
constraints on the absorbing cloud: the cloud has a column density of
$N_{\rm{H}} = 2.0^{+0.2}_{-0.3}\times 10^{22}$~cm$^{-2}$ which, once
combined with the estimated cloud density $n_{\rm c} \geq 1.5 \times
10^{9}$~cm$^{-3}$ implies a cloud size (diameter) $D_{\rm{c}}\leq 1.5
\times 10^{13}$~cm at a distance from the center of $R_{\rm c} \geq
4.3\times 10^{16}$~cm. Assuming that the cloud motion is dominated by
gravity the cloud (Keplerian) velocity is then $v_{\rm c} \leq
2100$~km~s$^{-1}$. All these properties are consistent with those of a
broad--line--region (BLR) cloud partially covering a compact X--ray
source during the {\it XMM--Newton} observation\footnote{The
  identification of the absorber with a BLR cloud is also supported by
  the observed optical broad line FWHM$\sim 2000$~km~s$^{-1}$ (Malizia
  et al. 2008). By assuming a flattened BLR geometry and the
  inclination derived from our disc reflection model ($i=44^\circ$),
  the FWHM translates into a Keplerian velocity of $1.4\times
  10^3$~km~s$^{-1}$, which is fully consistent with the cloud velocity
  upper limit $v_{\rm{c}} \leq 2.1 \times 10^3$~km~s$^{-1}$ we derive
  from our analysis.}.  The excellent agreement between the derived
cloud properties and the BLR physical conditions provides further
support to our proposed interpretation of the observed X--ray spectral
variability.

The partial eclipse can be used to constrain the size of the
X--ray emitting region. The $CF$ evolution timescales, as well as the
maximal observed covering fraction $CF_{\rm{max}}$ imply that the
cloud and X--ray emitting region sizes are related through $D_{\rm c}
= [0.66-1] \times D_{\rm s}$. The upper limit on the cloud size
$D_{\rm{c}}\leq 1.5\times 10^{13}$~cm then translates into an upper
limit on the X--ray source size of $D_{\rm s} \leq 2.3\times
10^{13}~{\rm{cm}} = 10.5~r_g/M_{\rm{best}}$ where $M_{\rm{best}}$ is the black hole mass in units of 
$1.5\times 10^7~M_\odot$, i.e. the best available estimate of the black
hole mass in \swi (Malizia et al. 2008). Such a small X--ray
emitting region is consistent with previous results based on similar
occultation events in the X--rays (Risaliti et al. 2007; 2009) as well
as with microlensing results (Dai et al. 2010; Mosquera et
al. 2013). A very compact X--ray corona is also consistent with the relatively
steep emissivity profile we (and Miniutti et al. 2009) measure for the
disc reflection component ($q\sim 5$) which strongly suggests a highly
compact X--ray emitting region (e.g. Miniutti \& Fabian 2004).

We conclude that we have observed the partial eclipse by a BLR cloud
of the X--ray continuum source in \swi. Our result is in line with the
mounting observational evidence that part of the observed X--ray
absorption in AGN is due to a clumpy absorber whose properties and
location can be identified with either the BLR or with a clumpy torus
(in those cases where absorption variability occurs on months to years
timescales) which probably are just different part of the same
obscuring region. Our analysis strongly suggests
that the X--ray continuum is produced in a compact region only a few
$r_g$ in size, which is consistent with the detection of a reflection
component off the inner accretion disc. Notice that the relatively
high observer inclination we measure from the broad Fe line ($i\sim
44^\circ$) likely enhances the probability of X--ray eclipses (see
e.g. Elitzur 2012). 

The advent of the next generation of large--collecting area X--ray
observatories (e.g. the proposed Athena+ mission, see Nandra et
al. 2013 and Dovciak et al. 2013) will allow us to map with even
greater accuracy the innermost regions of the accretion flow around
accreting black holes via X--ray spectroscopy and X--ray flux and
spectral variability studies, including absorption events as the one
discussed here (see e.g. Risaliti et al. 2011).

\section*{Acknowledgements}

Based on observations obtained with XMM-Newton, an ESA science mission
with instruments and contributions directly funded by ESA Member
States and NASA. Financial support for this work was provided by the
Spanish MINECO through grant AYA2010-21490-C02-02. The research
leading to these results has received funding from the European Union
Seventh Framework Programme (FP7/2007--2013) under grant n. 312789. MS
thanks CSIC for support through a JAE-Predoc grant. This work was
carried out despite the increasingly hostile cutbacks against the
Spanish system of science.

\end{document}